\documentclass[11pt,a4paper]{scrartcl}

\usepackage{ILD}

\usepackage[symbol]{footmisc}
\usepackage{feynmf}
\usepackage{multirow}
\usepackage{textpos}
\usepackage{amsmath}

\title{Measurement of the CPV Higgs mixing angle in ZZ-fusion at 1 TeV ILC}

\ildproc{phys}{2023}{006}

\date{\today}

\addauthor{N. Vuka\u{s}inovi\'{c}}{\institute{1}}
\addauthor{I. Bo\v{z}ovi\'{c}-Jelisav\u{c}i\'{c}}{\institute{1}}
\addauthor{G. Ka\u{c}arevi\'{c}}{\institute{1}}

\addinstitute{1}{``VIN\u{C}A'' Institute of Nuclear Sciences - National Institute of the Republic of Serbia, University of Belgrade, Mike Petrovi\'{c}a Alasa 12-14, 11351 Belgrade, Serbia}

\abstract{
Although the studies of tensor structure of the Higgs boson interactions with vector bosons and fermions at CMS and ATLAS experiments have established that the J$^\mathrm{PC}$ quantum numbers of the Higgs boson should be 0$^\mathrm{++}$, small CP violation in the Higgs sector (up to 10\% contribution of the CP-odd state) cannot be excluded with the current experimental precision. We review possibilities to measure CP violating mixing angle $\mathrm{\Psi_{CP}}$ between scalar and pseudoscalar states, at a linear electron-positron collider, at center-of-mass energy of 1 TeV.

\vspace{8cm}
\centering{Talk presented at\\ the International Workshop on Future Linear Colliders (LCWS 2023), 15-19 May 2023.} 
\newline
\centering{C23-05-15.3.}\\
\vspace{2cm}
\textit{This work was carried out in the framework of the ILD concept group}

}

\graphicspath{ {./logos/}{./figures/} }

\begin{document}

\titlepage

\section{Introduction}
\label{sec:intro}

In this study we present a method for the determination of the CP violating (CPV) mixing angle between scalar and pseudoscalar CP states. We assume that the SM-like Higgs boson 125 GeV mass eigenstate ($\mathrm{h_{125}}$) is a superposition of a scalar (CP-even) state H and a pseudoscalar (CP-odd) state A, via a mixing angle $\mathrm{\Psi_{CP}}$:

\begin{equation}
\label{hmix}
h_{125} = H \cdot cos\Psi_{CP} + A \cdot sin\Psi_{CP}
\end{equation}

A common framework for interpretation of CPV results from future Higgs $e^-e^+$ factories (both linear and circular), as well as from LHC and HL-LHC is given in the Snowmass CPV white paper \cite{rsnowm} for measurements based on angular observables and Effective Field Theory (EFT). The benchmark CPV parameter $f_{CP}$ quantifies contributions from CP-even and CP-odd amplitudes (here in example of a $H\rightarrow X$ decay) as:

\begin{equation}
\label{factorf}
f_{CP} = \frac{\Gamma_{H\rightarrow X}^{CP^{odd}}}{\Gamma_{H\rightarrow X}^{CP^{odd}} + \Gamma_{H\rightarrow X}^{CP^{even}}}
\end{equation}

\begin{figure}[h]
\includegraphics[width=\textwidth]{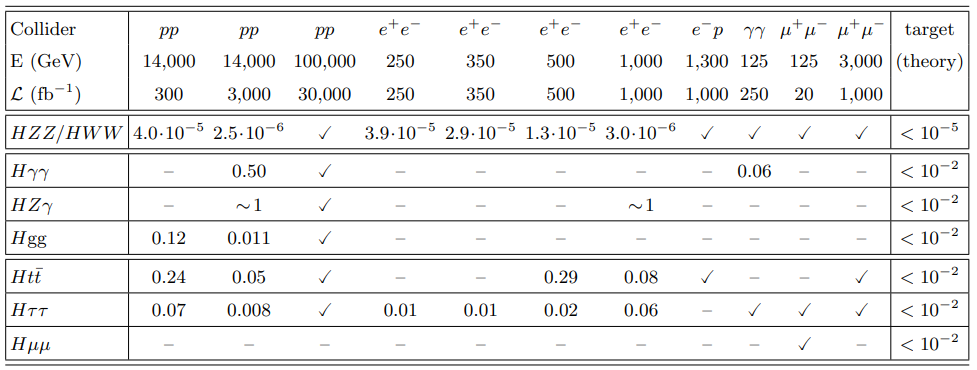}
\caption{Current estimates and theoretical target for $f_{CP}$ for different Higgs boson vertices at present and future colliders.}
\label{fig1}
\end{figure}

The sensitivity target for factor $f_{CP}$ is set from theory assuming 68\% CL measurement of the mixed Higgs state with up to 10\% CP-odd contribution. The estimates of $f_{CP}$ in the Higgs-vector boson vertices (HVV) are obtained from EFT theory in the Higgsstrahlung process. This is the first study addressing the $f_{CP}$ ($\Psi_{CP}$) determination in HVV vertices in Vector Boson Fusion. CPV in HVV vertices occurs at loop level, differently from Higgs-fermion vertices where it is realized at the Born level. 
As can be seen in Fig. \ref{fig1} \cite{rsnowm}, the precision target for factor $f_{CP}$ in HVV vertices is set to $< 10^{-5}$, assuming up to 10\% contribution of the CP-odd state.

The ILC (International Linear Collider) is a mature option for a linear $e^-e^+$ collider designed to operate from 250 GeV up to 500 GeV \cite{r2} in the center-of-mass, with projected integrated luminosities of 2 $\mathrm{ab^{-1}}$ and 4 $\mathrm{ab^{-1}}$, respectively. It offers possibility for upgrade to 1 TeV. Beam acceleration will be realized in superconducting accelerating cavities \cite{ilcacc}. The baseline design includes polarization scenario for $e^-$ and $e^+$ beams of 80\% and 30\%, respectively.
We have considered 1 $\mathrm{ab}^{-1}$ of data simulated either in the full ILD detector simulation (Mokka \cite{r3}) or fast ILD detector simulation (DELPHES \cite{r4}) with ILC-gen card, in ZZ-fusion at 1 TeV center-of-mass energy. Due to the fact that ZZ-fusion is a t-channel process $e^{\pm}$ final states are emmited at low polar angles. In the central tracker it will be found about 42\% of initially produced signal events. However, 1 TeV is found to be the optimal energy for this measurement in comparison to other center-of-mass energies, due to the most optimal interplay of centrality (which decreases with energy) and Higgs boson production cross-section (which increases with energy). As neutral current processes favor $e^-_{L}e^+_{R}$ and $e^-_{R}e^+_{L}$ polarisation of the initial state, measurements can be combined to reduce uncertainties. This is planned to be done, however at this instant we present results obtained with 100\% $e^-_{L}e^+_{R}$ beam polarization.

The ILD detector model \cite{rild} is assumed in this analysis. All detector subsystems are placed within a magnetic field of B = 3.5 T. ILD detector comprises: all-silicon vertex detector, gaseous (Time Projection Chamber) as a central tracker and compact Electromagnetic (ECAL) and Hadronic (HCAL) calorimeters. Apart from TPC, the tracking system comprises: two barrel components (the Silicon Inner Tracker SIT and the Silicon External Tracker SET), one End cap Tracker (ETD) and the Forward Tracker (FTD), covering between 140 mrad and 3000 mrad in polar angles. Excellent performances of the tracking system enable measurement of transverse momenta ($\mathrm{p_{T}}$) with an asymptotic resolution of $\sigma{ (1/p_{T})}$ $\sim$ 2 $\cdot$ 10$^{-5}$ GeV$^{-1}$ \cite{rild}. Particle reconstruction and identification relies on Particle Flow Algorithm (PFA) \cite{rpfa} using information from all detector subsystems. For example, PFA provides separation of jets that originate from Higgs boson and vector bosons ($\mathrm{W^{\pm}, Z^{0}}$) with 3-5\% jet energy resolution \cite{rild}.

\section{CPV sensitive observable}
\label{sec:zzf}
\begin{figure}[h]
\centering
\includegraphics[width=0.4\textwidth, height =0.4\textwidth  ]{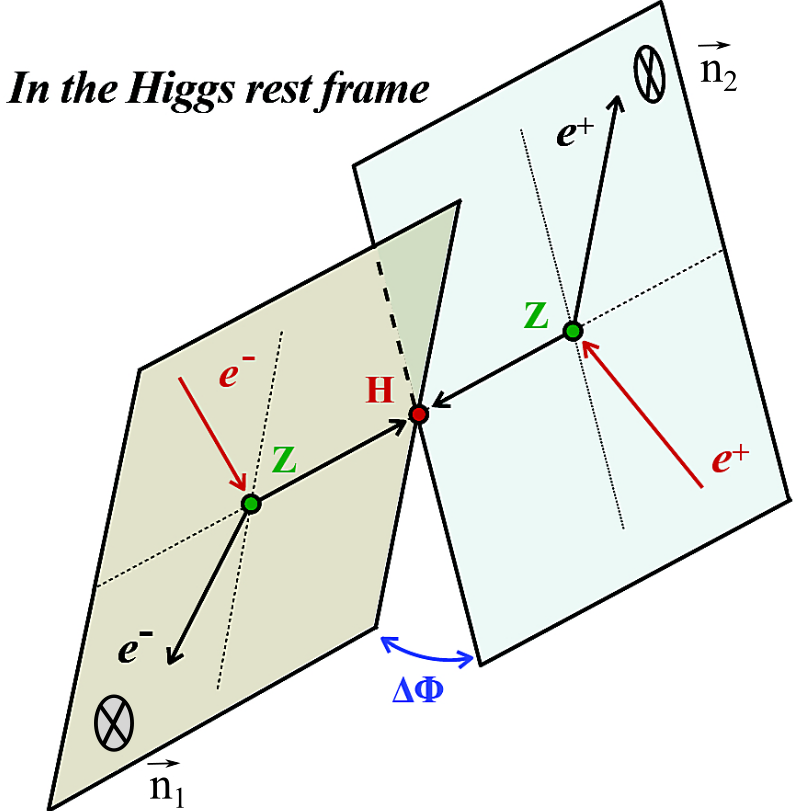}
\caption{Schematic view of production planes in ZZ-fusion and the angle $\Delta\Phi$ between them.}
\label{fig3}
\end{figure}

This analysis exploits the CPV-sensitive angular observable ($\Delta\Phi$), defined as the angle between production $e^-$ and $e^+$ planes in the Higgs rest frame, as illustrated in Fig. \ref{fig3}. One should note that the angle $\Delta\Phi$ is not the only CPV sensitive angular observable, but, as is shown in \cite{rsara}, it carries the most information on the CP nature of a Higgs boson.

$\Delta\Phi$ is calculated according to the following definition:

\begin{equation}
\label{deltafi}
\Delta\Phi =
\begin{cases}
      arccos(cos (\Delta\Phi)), \mathrm{sgn} (sin (\Delta\Phi)) \geq 0\\
      2 \pi - arccos(cos (\Delta\Phi)), \mathrm{sgn} (sin (\Delta\Phi)) \leq 0\\
        \end{cases}
\end{equation}

\noindent where 

\begin{equation}
\label{cosfi}
cos (\Delta\Phi) = \overrightarrow{n}_{1}\cdot\overrightarrow{n}_{2} \hspace{1cm} \mathrm{and} \hspace{1cm} \mathrm{sgn} (sin (\Delta\Phi)) = \frac{\overrightarrow{q}_{1}\cdot(\overrightarrow{n}_{1}\times\overrightarrow{n}_{2})}{|\overrightarrow{q}_{1}\cdot(\overrightarrow{n}_{1}\times\overrightarrow{n}_{2})|}
\end{equation}

\noindent Unit vectors $\overrightarrow{n}_{1}$ and $\overrightarrow{n}_{2}$ are defined by momenta of initial $\overrightarrow{q}_{{e}^{-(+)}_{i}}$ and final state $\overrightarrow{q}_{{e}^{-(+)}_{f}}$ electron (positron) and they are orthogonal to the corresponding production planes:

\begin{equation}
\label{n12}
\overrightarrow{n}_{1} = \frac{ q_{e^-_{i}}\times q_{e^-_{f}} }{ |q_{e^-_{i}}\times q_{e^-_{f}}| } \hspace{1cm} \mathrm{and} \hspace{1cm} \overrightarrow{n}_{2} = \frac{ q_{e^+_{i}}\times q_{e^+_{f}} }{ |q_{e^+_{i}}\times q_{e^+_{f}}|}
\end{equation}

\noindent In Eq. (\ref{n12}) $\overrightarrow{q}_{1}$ stands for momentum of $Z^{0}$ boson in the electron plane. Sign of the $sin (\Delta\Phi)$ from Eq. (\ref{cosfi}) actually shows if the positron plane rotates forward ($\mathrm{sgn}(sin (\Delta\Phi))$ = +1) or backward ($\mathrm{sgn}(sin (\Delta\Phi))$ = -1)  with respect to the electron plane following the right-handed rule with $Z^{0}$ boson in the electron plane emitted in the direction of a right-hand thumb.

\section{Event samples}
\label{sec:evtsamples}

The exclusive $H\rightarrow b\bar{b}$ final state is considered, to avoid high cross-section radiative processes ( $e^+e^- \rightarrow e^+e^-\gamma$). In case of signal event samples, about 3,500 events are fully simulated with the ILD detector and in the presence of the beam induced backgrounds. This statistics correspond to $\sim$ 1/5 of expected number of events in 1 $\mathrm{ab^{-1}}$ in full physical range of polar angles. About 28,000 events are simulated with Delphes V3.4.2 fast simulation, with ILD detector model. The relevant background (Table \ref{fig1}) is also fully simulated.

\begin{table}[!h]
\centering
\caption{\label{table:1} List of signal and background processes at 1 TeV with corresponding cross section $\sigma$, expected number of events in 1 ab$^{-1}$ $N_{evt}$, number of simulated events $N_{sim}$ and number of selected events in 1 ab$^{-1}$ of data $N_{sel}$. }
\begin{tabular}{ |l| l| l| p{3cm}| p{3 cm}|}
\hline
1 TeV & $\sigma (fb)$	& $N_{\mathrm{evt}} @ 1 \mathrm{ab^{-1}}$ & $N_{\mathrm{sim}}$ & $N_{\mathrm{sel}}@ 1 \mathrm{ab^{-1}}$   \\
\hline
Signal: $e^-e^+\rightarrow Hee; H\rightarrow b\bar{b}$ 	& 16 & 16016/8231$^{\mathrm{tracker}}$ & 27911 DELPHES, 3495 MC & 5658    \\ 
\hline
$e^-e^+\rightarrow q\bar{q}l^+l^- (l = \mu, \tau)^\dag$  & 255 & 255000 & 5886(1/43) & $/$      \\
\hline
$e^-e^+\rightarrow q\bar{q}$  & 9375 & 9375000 & 120343(1/78) & $/$   \\
\hline
$e^-e^+\rightarrow q\bar{q}l\nu^\dag$  & 4116 & 4116000 & 955058(1/4) & $/$    \\
\hline
\end{tabular}
\end{table}

\begin{footnotesize}
$\dag$ Currently there is only simulation for $l = \mu, \tau$. Electron state will be additionally added.
\end{footnotesize}

\begin{figure}[h]
\centering
\includegraphics[width=.4\textwidth, height=.35\textwidth]{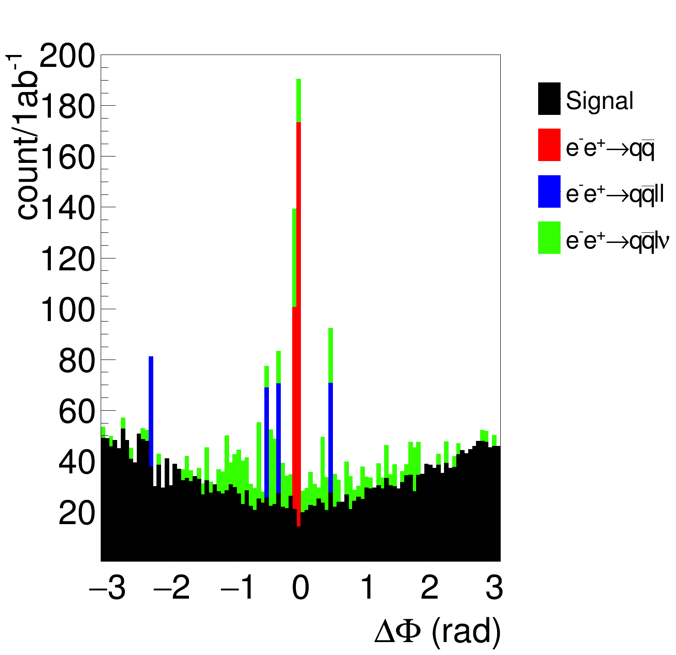}
\begin{textblock}{1.5}(2.3, -2.7)
\textit{\scriptsize ILD preliminary}
\end{textblock}
\hspace{10 mm} 
\includegraphics[width=.4\textwidth, height=.35\textwidth ]{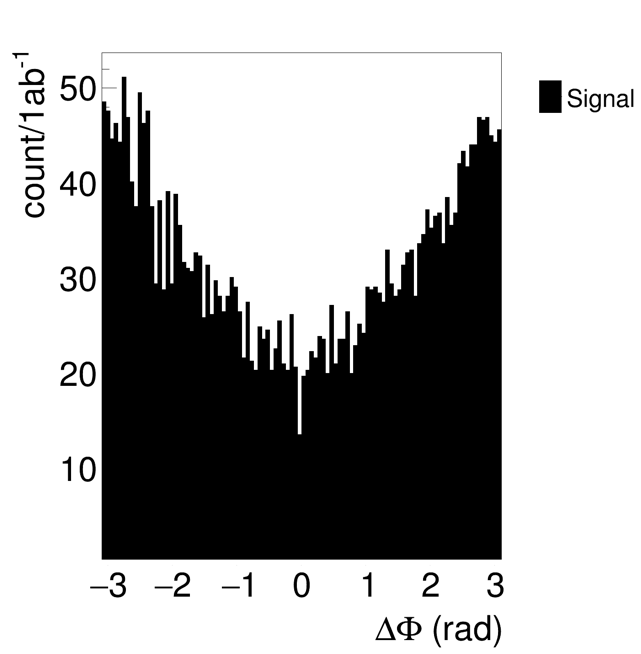}
\begin{textblock}{1.5}(8., -2.7)
\textit{\scriptsize ILD preliminary}
\end{textblock}
\caption{Stacked histogram of the $\Delta\Phi$ distribution for signal and background, after the preselection (left) and after the overall selection (right), for 1 $\mathrm{ab^{-1}}$ of data, in the tracking system of polar angles.}
\label{figstack}
\end{figure}

\begin{figure}[h]
\centering
\includegraphics[width=.45\textwidth, height=.35\textwidth]{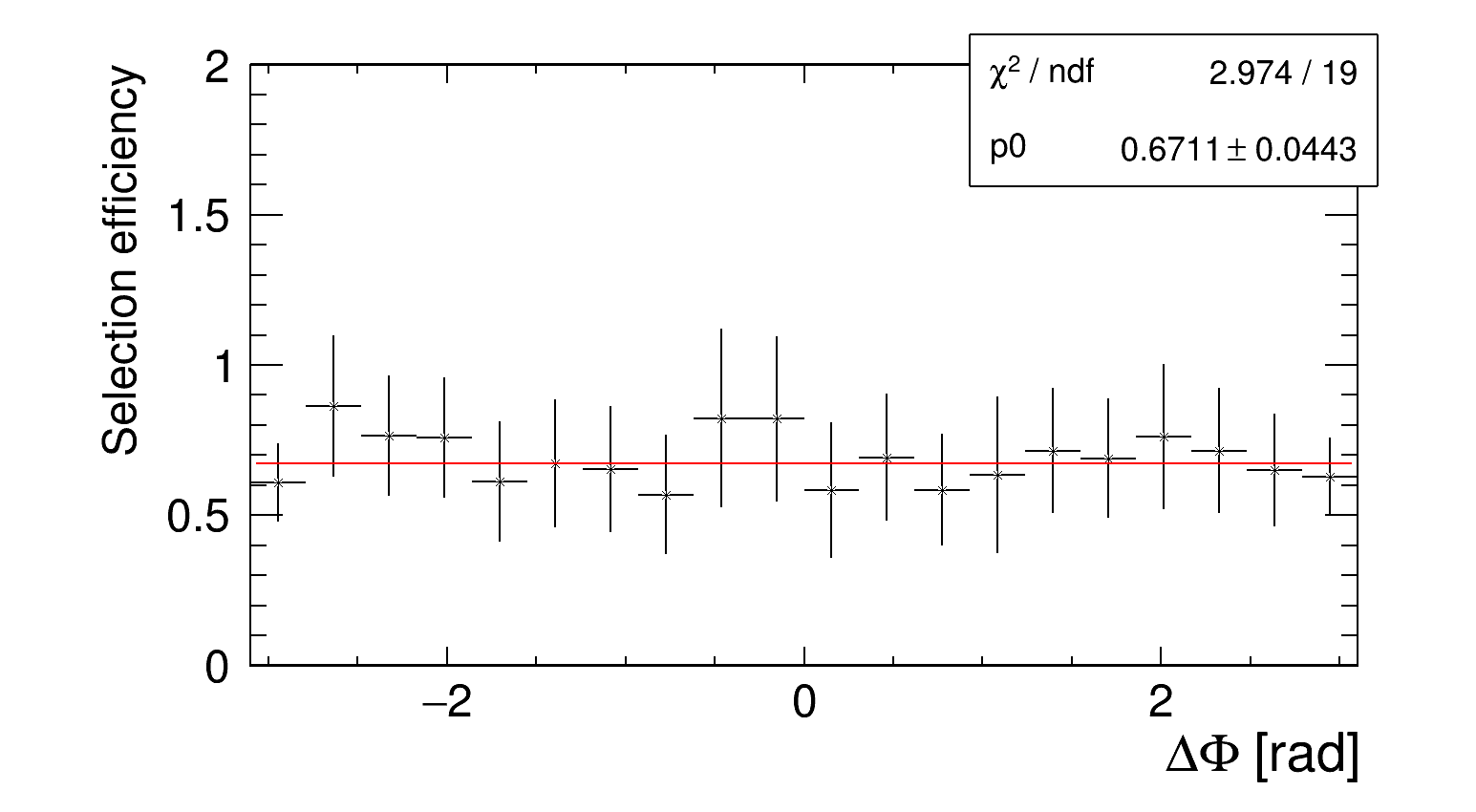}
\hspace{5 mm} 
\includegraphics[width=.45\textwidth, height=.35\textwidth ]{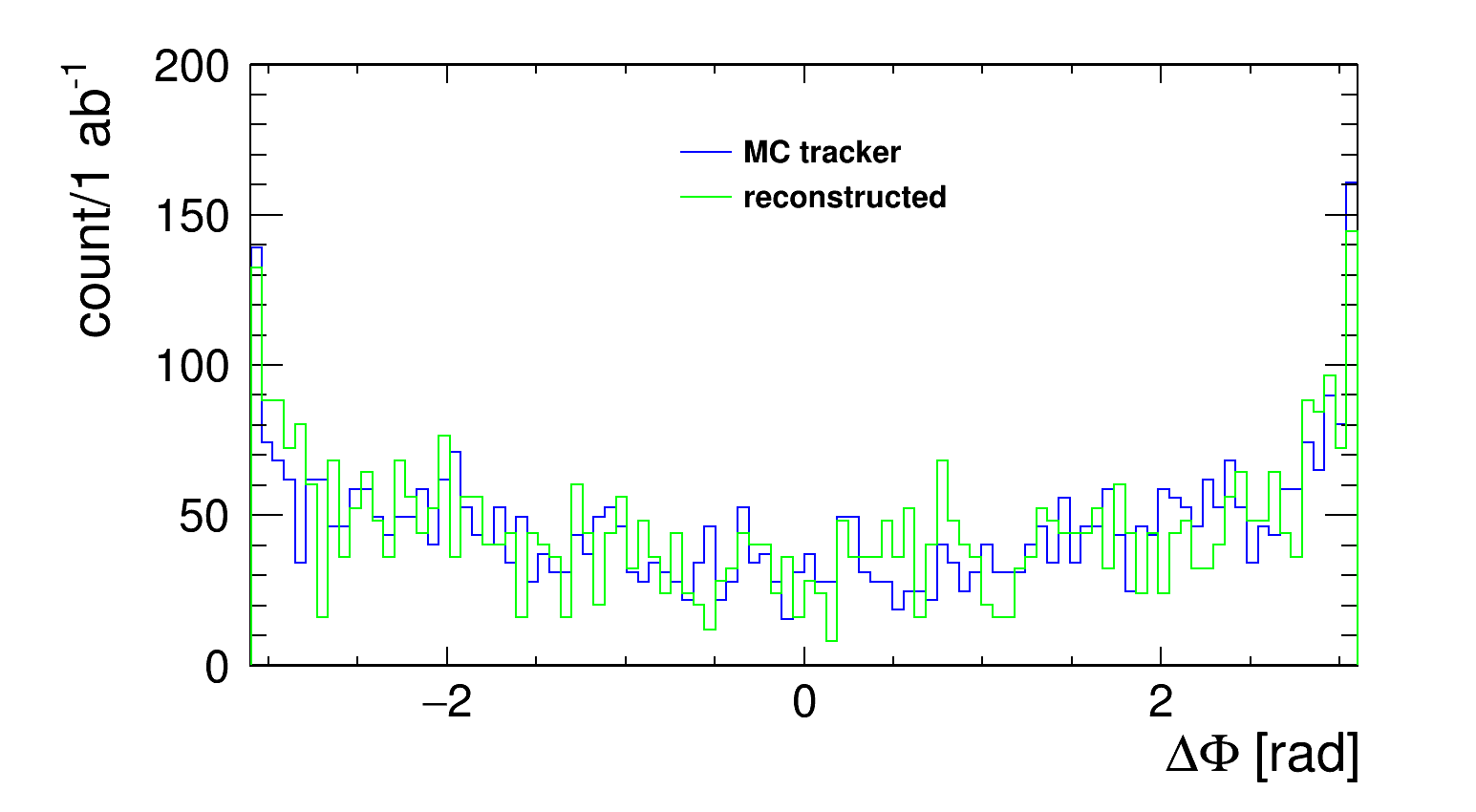}
\begin{textblock}{1.5}(8.5, -2.7)
\textit{\scriptsize ILD preliminary}
\end{textblock}
\caption{Unbiased event selection for the pure scalar state (left). $\Delta\Phi$ distributions from the reconstructed data (green) and at the generator level (blue), where the reconstruction is performed with Delphes (right). }
\label{figfios}
\end{figure}

\begin{figure}[h]
\centering
\includegraphics[width=.45\textwidth, height=.35\textwidth]{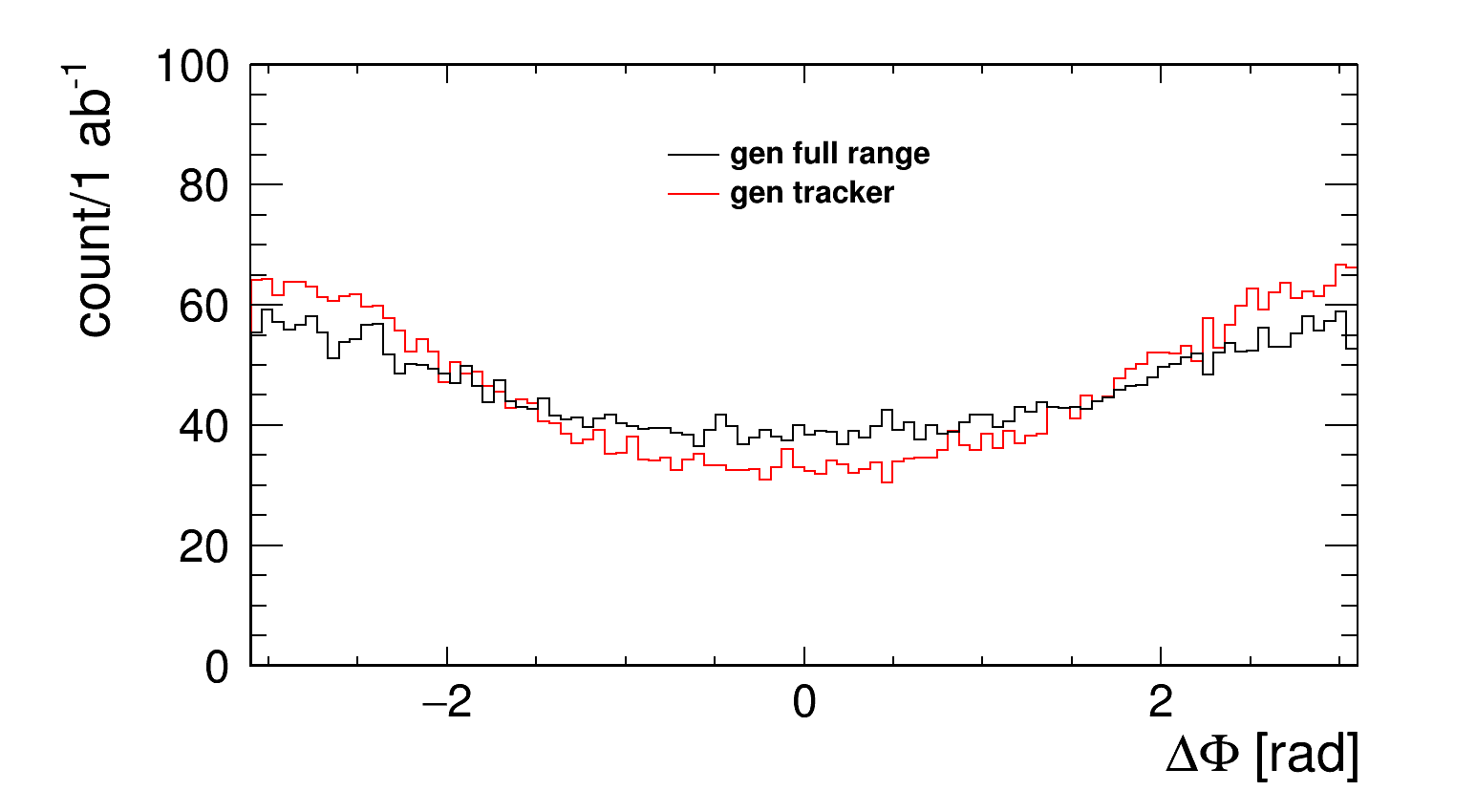}
\begin{textblock}{1.5}(1.5, -2.7)
\textit{\scriptsize ILD preliminary}
\end{textblock}
\hspace{5 mm} 
\includegraphics[width=.45\textwidth, height=.35\textwidth ]{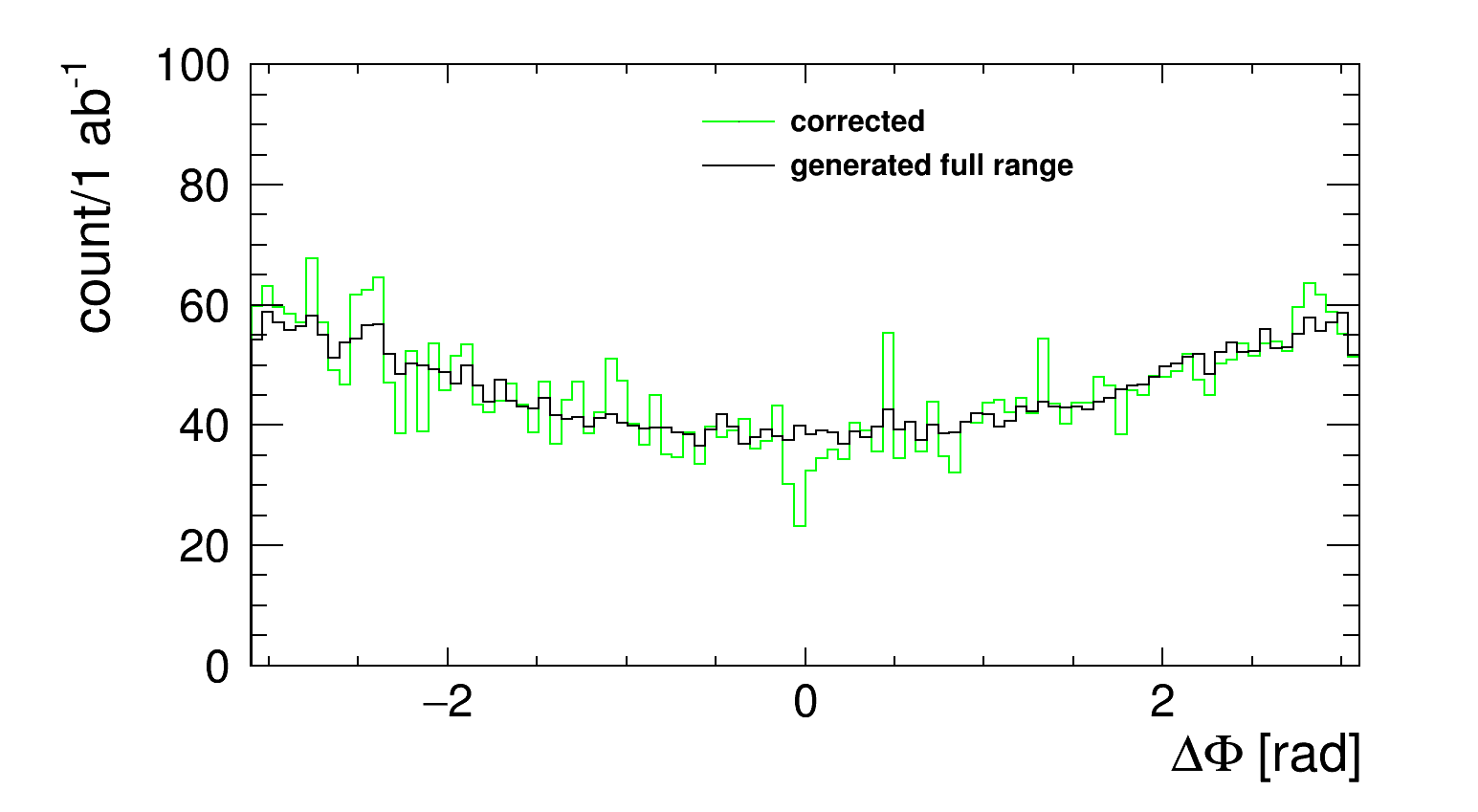}
\begin{textblock}{1.5}(7.5, -2.7)
\textit{\scriptsize ILD preliminary}
\end{textblock}
\caption{$\Delta\Phi$ distribution at the generator level in full physical range (black line) and in the tracker (red line) (left). Comparison of $\Delta\Phi$ distributions at the generator level in full physical range (black line) and after correction for the central tracker acceptance effects (green line) (right). }
\label{figgencor}
\end{figure}

By applying electron isolation in Delphes, events with exactly one isolated electron and one isolated positron are preselected. Rest of the particles are clustered in 2 jets by the Durham algorithm \cite{rdurham}. After preselection, about 1170 background events are left in 1 $\mathrm{ab^{-1}}$ of data (Fig. \ref{figstack} (left)). In general, $\Delta\Phi$ distribution for background is flat as it is CP-insensitive, though it is not obvious from Fig. \ref{figstack} (left) where limitted number of background events is selected with large scaling factors.
Further event selection requires: cut on di-jet invariant mass: 80 GeV $< m_{q\bar{q}} <$ 160 GeV, cut on Z boson masses: $m_{Z_{1}, Z_{2} } >$ 30 GeV, cut on transverse momentum of final state system electron and positron: $p_{T_{ee}} >$ 15 GeV, cut on missing transverse momentum: $p_{T_{miss}}<$ 150 GeV. After applying all these criteria background is fully suppressed while the total signal efficiency, including preselection, is about 68\%. $\Delta\Phi$ for signal after full event selection is given in the Fig. \ref{figstack} (right). Event selection doesn't bias $\Delta\Phi$ distribution and that is illustrated in the Fig. \ref{figfios} (left).

In order to quantify the impact of polar angle acceptance and detector reconstruction effects on the $\Delta\Phi$ observable, Whizard generator version 2.8.3 \cite{rwhizard} is used with the Higgs characterization model \cite{r8} within the UFO framework. As can be seen from Fig. \ref{figgencor} (left), acceptance limited by the tracker polar angles affects the $\Delta\Phi$ distribution with respect to the full physical range. On the other hand, comparing generated and reconstructed information in the tracker, it is obvious that detector effects are negligible as is illustrated in the Fig. \ref{figfios} (right). Thus reconstructed information has to be corrected for the acceptance effects. Corrected $\Delta\Phi$ distribution is illustrated in Fig. \ref{figgencor} (right). It reasonably reproduces distribution at the generator level in the full physical range, up to statistical fluctuations.

\section{$\mathrm{\Psi_{CP}}$ determination}

\begin{figure}[h]
\centering
\includegraphics[width=.45\textwidth, height=.35\textwidth]{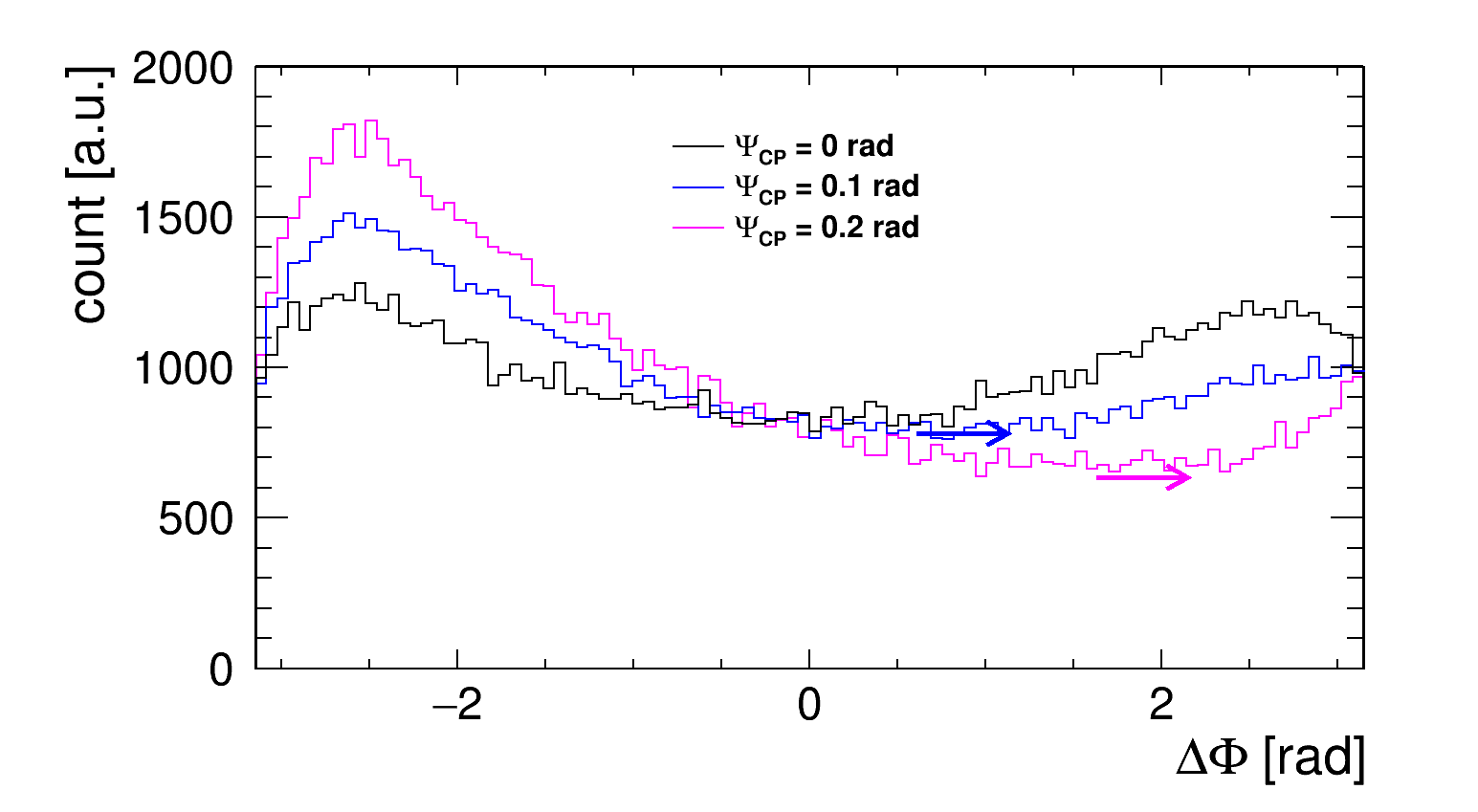}
\hspace{5 mm} 
\includegraphics[width=.45\textwidth, height=.35\textwidth ]{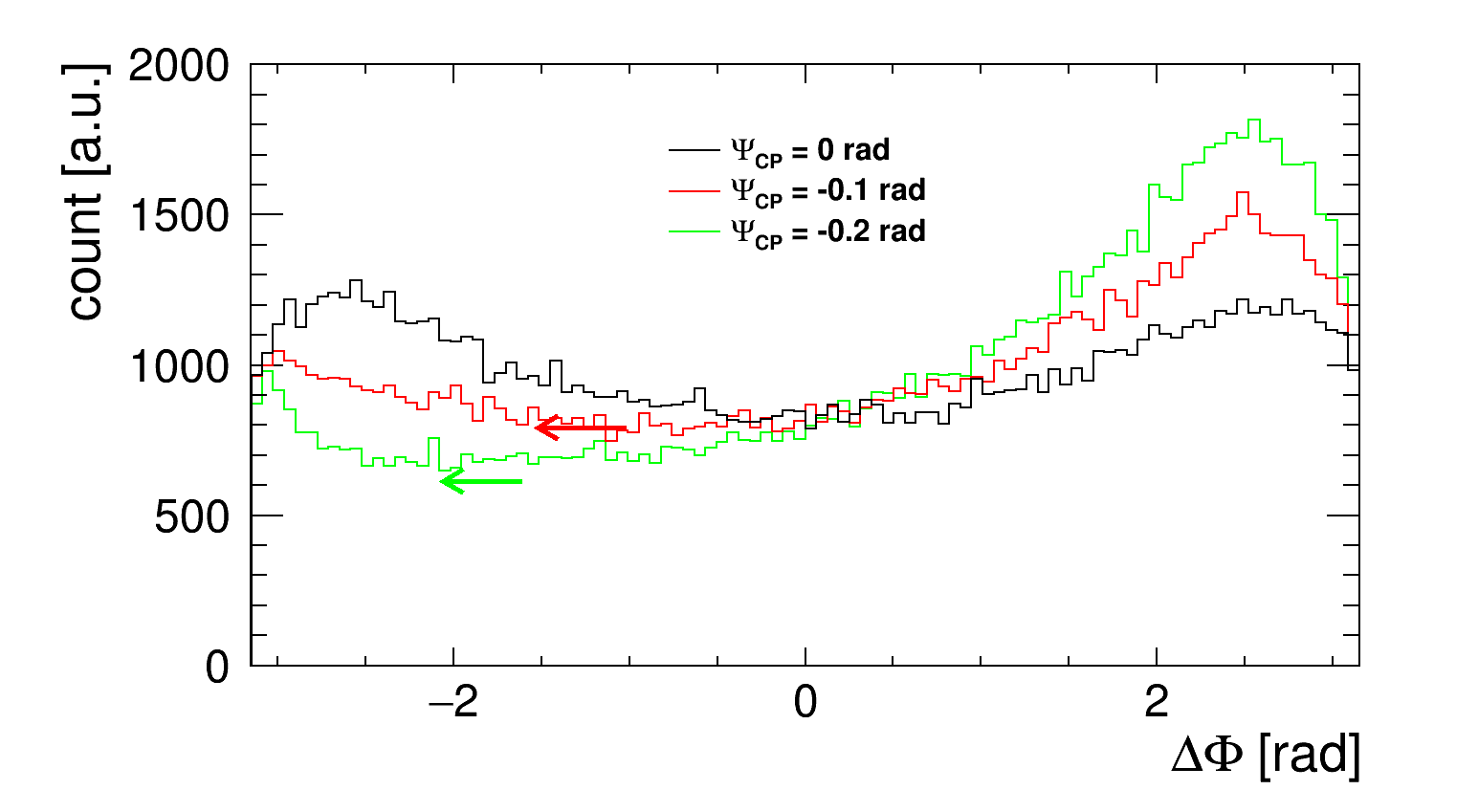}
\caption{$\Delta\Phi$ distribution for negative $\mathrm{\Psi_{CP}}$ values (left) and positive $\mathrm{\Psi_{CP}}$ values (right), illustrates the corresponding shift of the local minimum.   }
\label{figfipsi}
\end{figure}

Unlike $H\rightarrow \tau^+\tau^-$ decays where the shape of the angular observable is derived from the differential cross-section \cite{rjeans}, dependence of $\Delta\Phi$ on CP violating mixing angle $\mathrm{\Psi_{CP}}$ is not known. Due to that we applied a phenomenological approach in extraction of $\mathrm{\Psi_{CP}}$ from simulated (or experimental) data. We have assumed CP-odd admixture up to 17\%, corresponding to $\mathrm{\Psi_{CP}}$ values up to 0.2 rad. $\Delta\Phi$ distributions are changing for different assumptions on $\mathrm{\Psi_{CP}}$ value, in a way that the local minimum gets shifted accordingly (to the left or right for negative or positive $\mathrm{\Psi_{CP}}$ values, respectively). This is illustrated in Fig. \ref{figfipsi} (left and right)).

\begin{figure}[h]
\centering
\includegraphics[width=.45\textwidth, height=.35\textwidth]{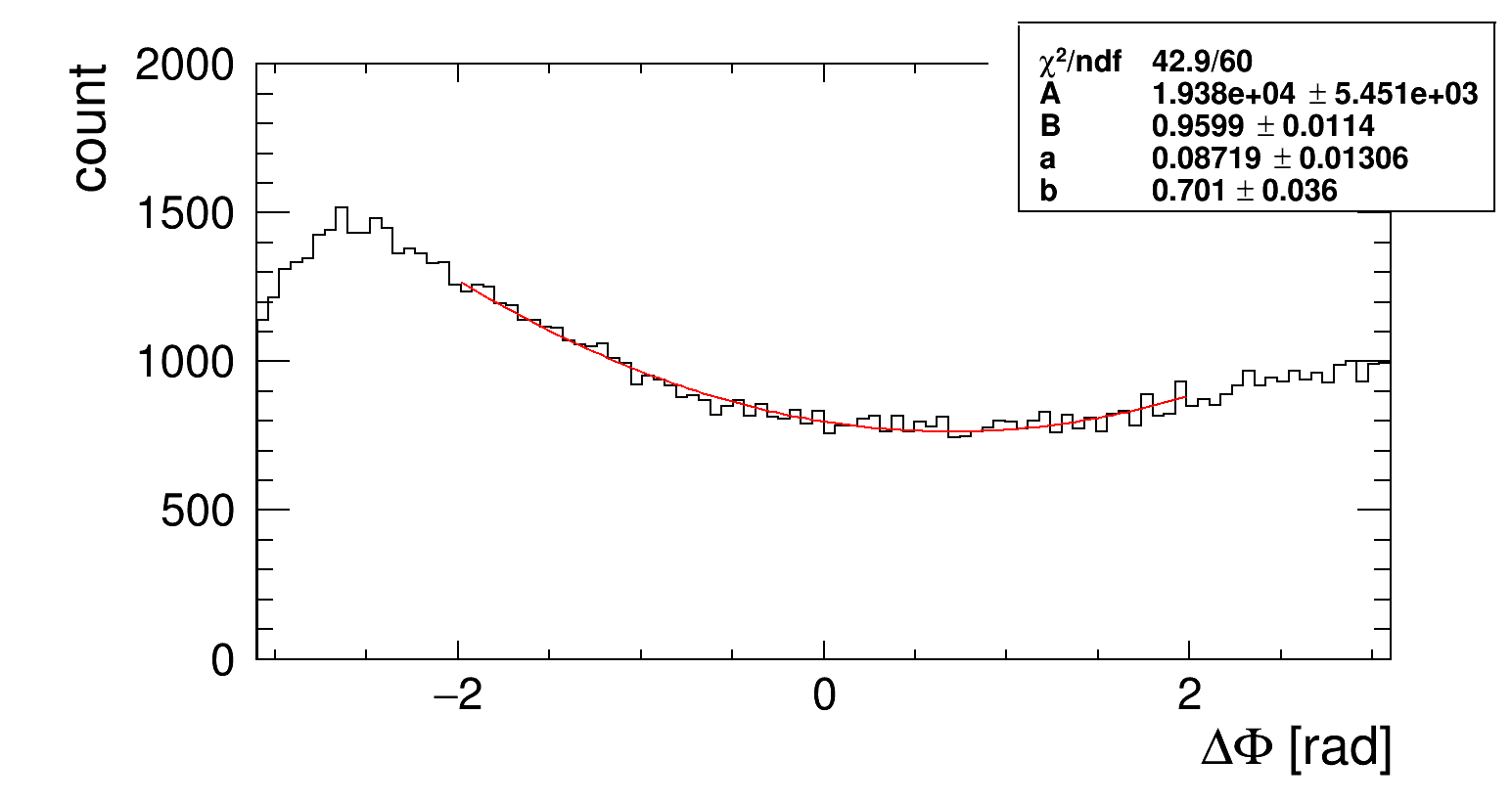}
\hspace{5 mm} 
\includegraphics[width=.45\textwidth, height=.35\textwidth ]{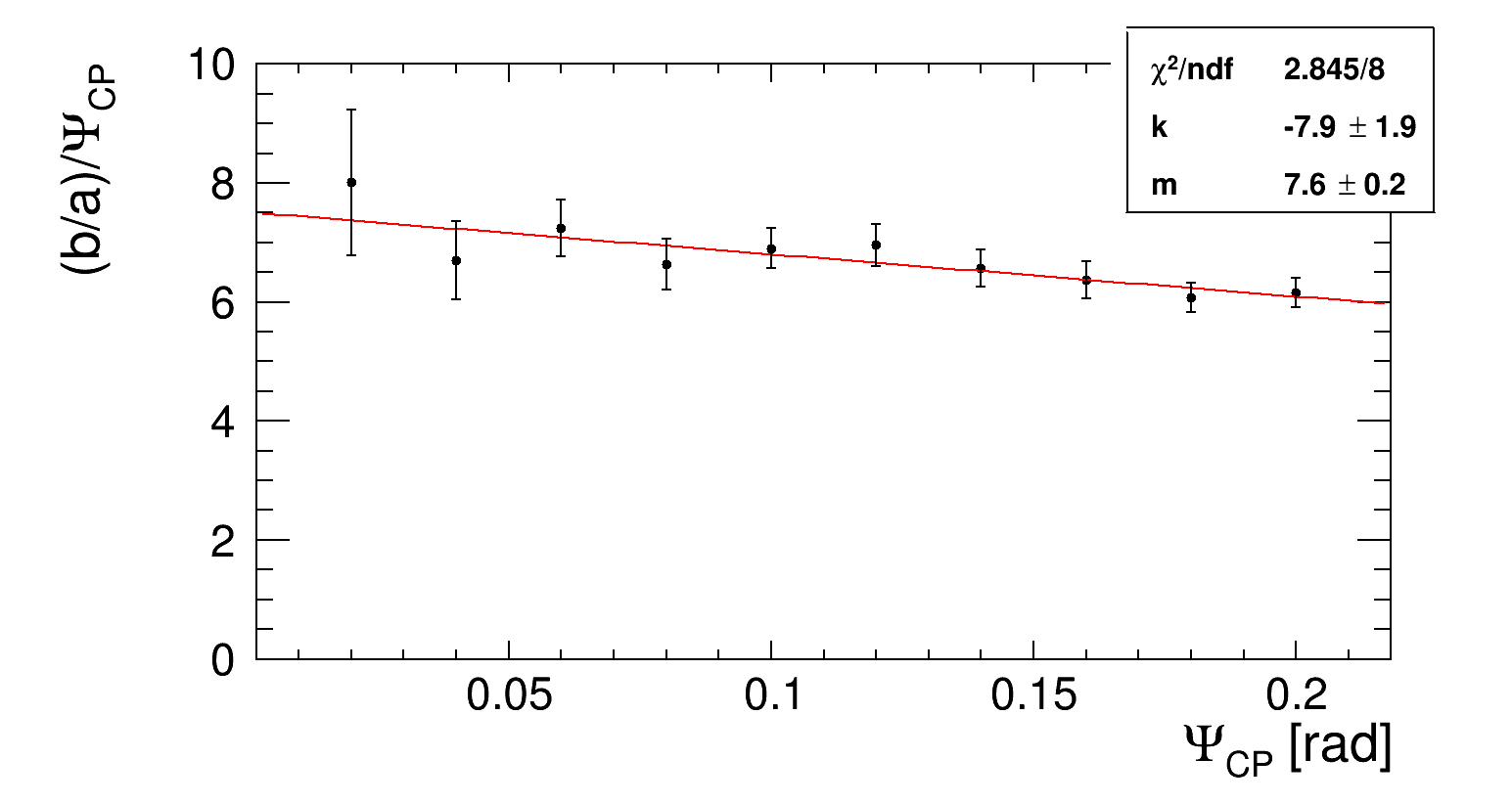}
\caption{Fit of $\Delta\Phi$ distribution around local minimum for $\mathrm{\Psi_{CP}}$ = 0.1 in order to determine position of the minimum (left). Dependence of the local minimum position $b/a$ on true $\mathrm{\Psi_{CP}}$ values (right). }
\label{figkoefkm}
\end{figure}

We perform a fit around local minimum with the fit function to determine its position:

\begin{equation}
\label{fitfunc}
f(\Delta\Phi, \mathrm{\Psi_{CP}}) = A + B \cdot cos(a \cdot \Delta\Phi - b)     
\end{equation}

\noindent where $\mathrm{A, B}, a$ and $b$ are free parameters, and the ratio of coefficients $b$ and $a$ gives the minimum of the $\Delta\Phi$ distribution.
Example of such a fit performed on 10${^5}$ generated events for $\mathrm{\Psi_{CP}}$ = 0.1 is illustrated in Fig. \ref{figkoefkm} (left). Position of the minimum ($b/a$) over $\mathrm{\Psi_{CP}}$ true value is a linear function of the true value of $\mathrm{\Psi_{CP}}$. This is illustrated in Fig. \ref{figkoefkm} (right). 

Thus the following equation holds:
\begin{equation}
\label{bia}
(b/a)/\mathrm{\Psi_{CP}} = k \cdot \mathrm{\Psi_{CP}} + m     
\end{equation}

\begin{figure}[h]
\centering
\includegraphics[width=.5\textwidth, height=.35\textwidth]{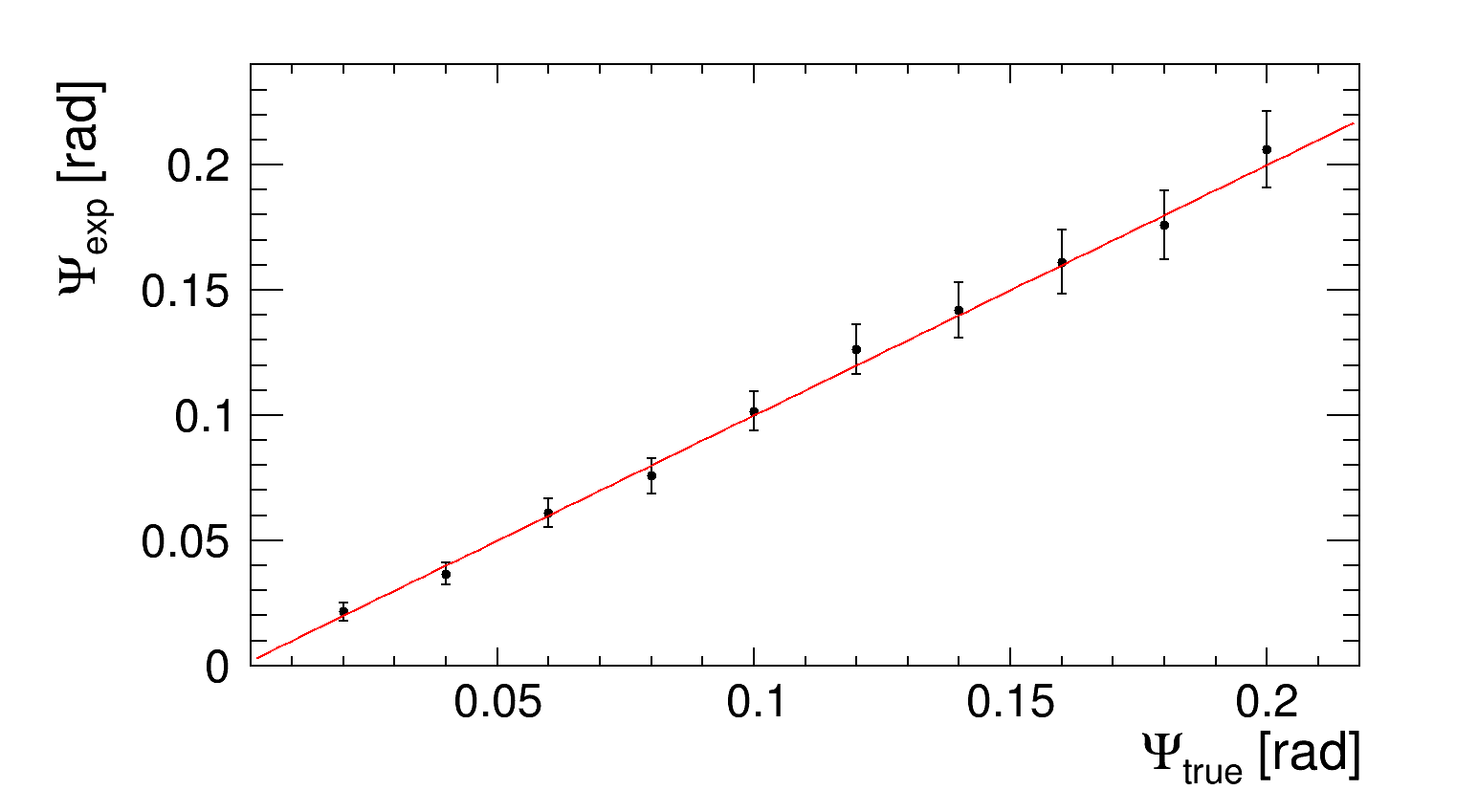}
\caption{Dependence of extracted $\mathrm{\Psi_{exp}}$ values with respect to the true ones ($\mathrm{\Psi_{true}}$).}
\label{dissip}
\end{figure}

\noindent Coefficients $k$ and $m$ can be determined from simulation assuming certain range of $\mathrm{\Psi_{CP}}$ values, while minimum can be measured from the (experimental) data as is illustrated in Fig. \ref{figkoefkm} (left). Thus $\mathrm{\Psi_{CP}}$ can be determined from Eq. (\ref{bia}) by solving this quadratic equation. Dependence of the extracted values $\mathrm{\Psi_{exp}}$ versus the true values $\mathrm{\Psi_{true}}$ is given in Fig. \ref{dissip}. For the $\mathrm{\Psi_{true}} \leq 0.1$ corresponding to up to 9\% admixture of the CP-odd state differences between $\mathrm{\Psi_{true}}$ and $\mathrm{\Psi_{exp}}$ central values is not larger than 5 mrad. 

\section{$\mathrm{\Psi_{CP}}$ from reconstructed data - a pseudo-experiment} 

\begin{figure}[h]
\centering
\includegraphics[width=.5\textwidth, height=.4\textwidth]{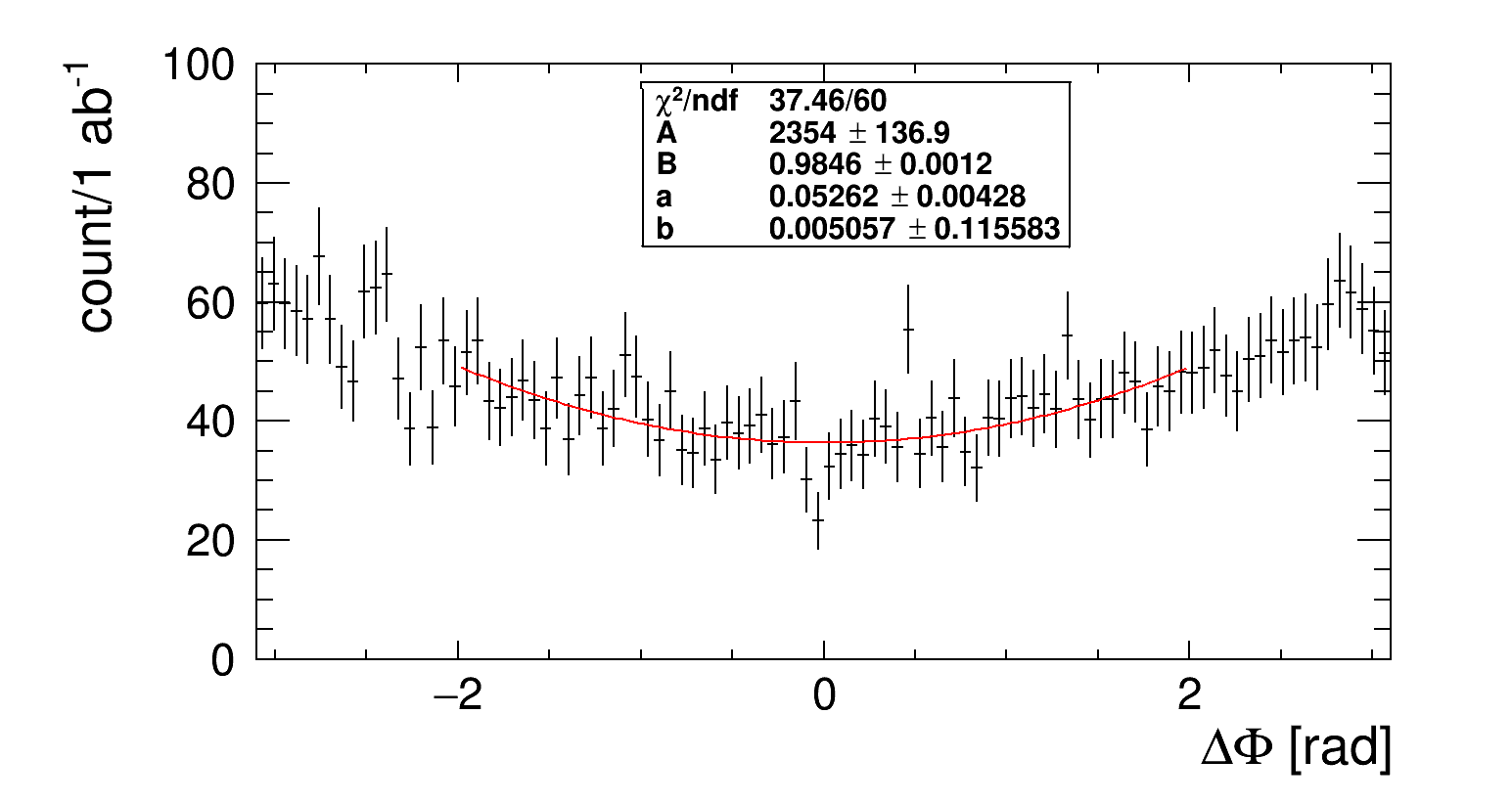}
\caption{Fit of $\Delta\Phi$ local minimum on the corrected reconstructed data for pure scalar state.}
\label{phireconstr}
\end{figure}

\begin{figure}[h]
\centering
\includegraphics[width=.45\textwidth, height=.35\textwidth]{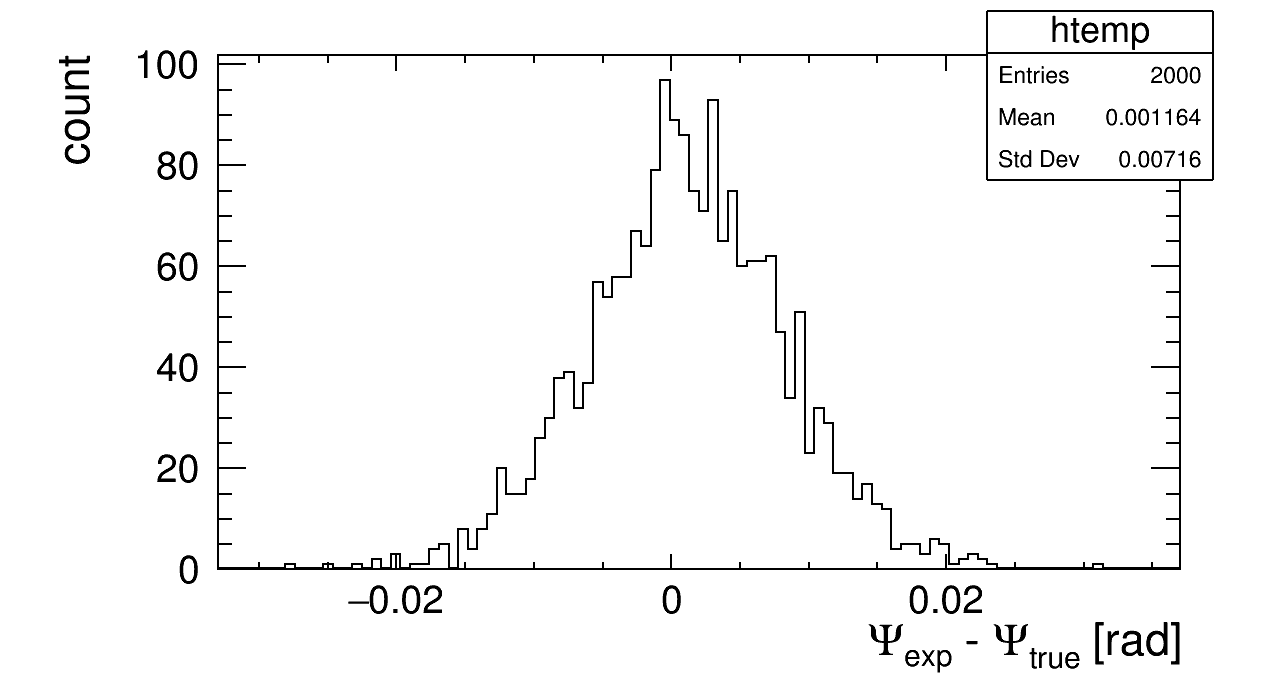}
\caption{Statistical dissipation of the extracted $\mathrm{\Psi_{CP}}$ value with respect to the $\mathrm{\Psi_{true}}$ = 0, in 2,000 repeated pseudo-experiments each with 1 $\mathrm{ab^{-1}}$ of 100\% $e^-_{L}e^+_{R}$ polarized data. }
\label{pseudo}
\end{figure}

Using the parameters ($k$ and $m$) of the linear fit from simulation (Fig. \ref{figkoefkm} (right)), measurement of the local minimum from the reconstructed data (Fig. \ref{phireconstr}) gives the value of $\mathrm{\Psi_{CP}}$ = 0.9 mrad for the pure scalar state. In order to estimate statistical dissipation of the extracted $\mathrm{\Psi_{CP}}$, we run 2000 pseudo-experiments for the pure scalar state, each experiment with about 1 $\mathrm{ab^{-1}}$ of generated data. This is illustrated in Fig. \ref{pseudo}. Absolute statistical uncertainty ($\Delta\mathrm{\Psi_{CP}}$) of extracted $\mathrm{\Psi_{CP}}$ value is determined by RMS of the distribution from Fig. \ref{pseudo} reflecting the probabilistic nature of the analized sample in the context of statistical population. RMS is found to be 7 mrad. $\Delta\mathrm{\Psi_{CP}}$ value of 7 mrad corresponds to 68\% CL uncertainty to measure $\mathrm{\Psi_{CP}}$ = 0 in 1 ab$^{-1}$ of data. The latter translates to uncertainty of the CPV factor $f_{CP} \approx$ 4.9 $\cdot$ 10$^{-5}$, that is still slightly above the theoretical target. Further improvement is expected in combination of results obtained on samples with different beam polarizations, as mentioned in Sec.\ref{sec:intro}. Systematic uncertainty from the fit $\leq$ 1 mrad is significantly smaller compared to the statistical one.

\section{Conclusion}
In this paper we present the result of the first CPV mixing angle measurement based on angular observable, in HVV vertex at an $e^-e^+$ collider. Measurement is simulated in ZZ-fusion at 1 TeV ILC with the ILD detector. Local minimum of the angular observable is sensitive to the  $\mathrm{\Psi_{CP}}$  mixing angle and can be obtained from the phenomenological fit of the reconstructed (simulated or experimental) data. Individual measurement on the fully simulated data gives deviation of 0.9 mrad from the truth value for the pure scalar state. The method is stable for $\mathrm{\Psi_{CP}}$  variations up to 0.2 rad, corresponding to pseudoscalar admixture of $\sim$ 17\%. At 1 TeV ILC with 1 ab$^{-1}$ of $e^-_{L}e^+_{R}$ data, pure scalar should be measured with statistical uncertainty of 7 mrad at 68\% CL. Obtained statistical uncertainty corresponds to the precision of the CPV factor $f_{CP} \approx$ 4.9 $\cdot$ 10$^{-5}$. The obtained precision is not yet final, as ILC offers possibility to combine different beam polarizations.

\section*{Acknowledgements}
We would like to acknowledge our colleagues from ILC IDT Working Group 3 for the technical support as well as for comments on the physics content, especially to Aleksander Filip \.{Z}arnecki for useful ideas and discussions leading to better understanding of the sensitive observable behavior.   
This research was funded by the Ministry of Education, Science and Technological Development of the Republic of Serbia and by the Science Fund of the Republic of Serbia through the Grant No. 7699827, IDEJE HIGHTONE-P.


\end{document}